\documentclass[conference]{IEEEtran}
\IEEEoverridecommandlockouts
\usepackage{cite}
\usepackage{amsmath,amssymb,amsfonts}
\usepackage{iftex}
\ifPDFTeX
  \usepackage{times}
\else
  \usepackage{fontspec}
\fi
\usepackage{graphicx}
\usepackage{textcomp}
\usepackage{xcolor}
\usepackage{tikz}
\usetikzlibrary{arrows.meta,positioning}
\begin{document}

\title{A Scenario-Knowledge-Driven Pipeline\\ for Just-in-Time Assistance}

\author{\IEEEauthorblockN{Zhiyuan Li, Tatsunori Hara, and Jun Ota}
\IEEEauthorblockA{\textit{Research into Artifacts, Center for Engineering (RACE)} \\
\textit{Graduate School of Engineering, The University of Tokyo}, Tokyo, Japan \\
\{zhiyuan430602, hara-tatsunori\}@g.ecc.u-tokyo.ac.jp, ota@race.t.u-tokyo.ac.jp}
\thanks{The 4th Workshop on Nonverbal Cues for Human--Robot Cooperative
Intelligence, IEEE/RSJ International Conference on Intelligent Robots and
Systems (IROS 2026), October 1, 2026, David L. Lawrence Convention Center,
Pittsburgh, PA, USA.}
}

\maketitle

\begin{abstract}
Detecting a silently struggling kiosk user is only the first step; deciding
whether, when, and how to help depends on scenario knowledge usually buried
in model weights and thresholds. We
propose a scenario-knowledge-driven pipeline: a single scenario knowledge
document, human-authored and version-controlled, configures sensing, constrains LLM
reasoning, and shapes a graded intervention proposal. Narration,
assistance-need assessment, and proposal are kept separate for independent
audit. As proof of concept, we replay two recorded kiosk sessions offline, chosen before the
runs for their struggle evidence and retrospective detail. Both cases support what the design
promises: checkable reporting and measured
escalation. Across 95 updates, every sentence of the append-only narration
cites the primitive events underlying it, and the rule layer detects 12 of 13
and 7 of 7 annotated struggle episodes under a strict criterion. The assessor
de-escalates on recovery and reaches the top rung exactly once, under
maximally converging evidence. At the decisive help-seeking turn, narration,
assessment, and the participants' retrospective accounts converge. The appropriateness of these interventions, the
pipeline's restraint on sessions without struggle, and the document's transfer
to a new scenario frame the agenda.
\end{abstract}

\begin{IEEEkeywords}
nonverbal cues, proactive assistance, large language models, human--robot cooperative intelligence
\end{IEEEkeywords}

\section{Introduction}\label{sec:intro}
Public self-service kiosks (ticketing machines, transit terminals, government
service counters) assume users can complete their task unaided. Many cannot,
and a large share of those who struggle never explicitly ask for help:
they hesitate, repeat the same step, glance around for a staff member, sometimes
abandon the task, all without a word. We call these users \emph{silent
strugglers}. Recognizing them before they give up enables just-in-time
assistance: support at the moment of need and receptivity, not after visible
failure or on a fixed schedule \cite{nahum2018jitai}.

Our prior project, JITAS, asked whether silent struggling at a public ticket
kiosk is visible to a machine before the user says anything. From synchronized
multimodal recordings, we reported that help-seeking intention is signaled by
the \emph{temporal variability} of facial expression, the way the face changes
over time rather than any single configuration \cite{li2026jitas}. The nonverbal signal
exists and a machine can read it; that alone does not tell a system what to do.

Detection is only the first step. Knowing that a user may need help
does not decide whether, when, or how to act. A poorly timed or overbearing
offer can be as harmful as none. It can interrupt a coping user or erode
the autonomy of one who is not. What makes these decisions workable in
deployment is scenario-specific knowledge fixed at design time: task structure,
user population, salient signals, plausible assistance-need states, available
interventions, what counts as socially acceptable help. Each assumption may be
sensible; the trouble is it lives \emph{implicitly}, across weights, thresholds,
and hard-coded branches, beyond inspection, review, or transfer.

This paper asks what becomes possible when that knowledge is explicit. We
propose a scenario-knowledge-driven pipeline in which a single architecture reads
a version-controlled \emph{scenario knowledge document} and uses
it to configure sensing, constrain LLM reasoning, and shape intervention
selection. The document is human-authored and inspectable; it specifies task
structure, observable signals, plausible assistance-need states, graded
intervention options, and social norms. By design, the system keeps three outputs distinct: an incrementally
updated observation narrative, an assistance-need assessment, and a graded
intervention proposal, so that every inference step can be audited on its own
terms. Our contributions are: (i) a framework that treats scenario knowledge as
an explicit, replaceable configuration for adaptive assistance from nonverbal
cues; (ii) a first instantiation as a multi-timescale pipeline built
around that document as its central artifact; and (iii) a preliminary kiosk case
study in which each observation sentence cites the primitive events underlying it,
demonstrating the detection-to-decision chain end to end. The \emph{appropriateness} of a chosen intervention is beyond this
paper's scope.

\section{Related Work}\label{sec:related}
\textbf{Assistance-need from nonverbal cues.} In intelligent tutoring, facial
action units and head pose estimate engagement and help-seeking so hints arrive
on time \cite{wang2025helpseek}. In socially assistive robotics, gaze fused with
language cues decides when to assist with minimal task modeling
\cite{wilson2021socialcues}. Our own prior work detects help-seeking intention
at a public ticket kiosk from the temporal variability of facial expression
\cite{li2026jitas}. These efforts share the logic of just-in-time adaptive
intervention \cite{nahum2018jitai}. Most, however, stop at detection, leaving open
how to act on an uncertain need estimate under scenario-specific constraints.

\textbf{LLM- and VLM-driven proactive interaction.} A second line uses large
language and vision-language models to decide when and how to act. Closest
in spirit, Attentive Support's LLM-based robot decides when to help a group and
when to stay silent to avoid disturbing it \cite{tanneberg2024help}. LaMI
regulates multimodal robot behavior through natural-language guidance instead
of hand-built state machines \cite{wang2024lami}. Our scenario knowledge document also selects which detectors run,
bounds the inference space, and is frozen under version control. The document, not the code, is thus what a new scenario replaces. MERGE gates an
expensive vision-language model behind a lightweight change detector
\cite{deigmoeller2026merge}; our instantiation
(Section~\ref{sec:framework}) gates LLM calls behind rule-emitted events in the
same spirit. Further mechanisms include distilled vision-language knowledge for
proactive utterances \cite{desimone2025proactive}, procedure tracking that
contextualizes assistance \cite{arakawa2025prism}, streaming narration of
egocentric video into dialogue \cite{zhang2025proassist}, on-demand sensory
context \cite{yang2025proagent}, and sensor-language representations
\cite{zhang2025sensorlm}. None, however, offers a reusable public-service
stack or treats scenario knowledge as an explicit, inspectable, swappable
artifact.

\textbf{Perceptual primitives and appropriateness.} Task-specific perceptual
primitives (reusable detectors) can be selected from a pool; end-to-end
gaze-target detection suits kiosk settings \cite{lin2025gazehta}. Beyond accuracy, proactive assistance
raises appropriateness questions: unsolicited help can undermine agency or
become surveillance, and reasonable deviation can be misread as error
\cite{glawe2025agency}. These
concerns motivate graded, auditable, explicitly configured intervention rather
than maximal proactivity.

\section{Scenario-Knowledge-Driven Pipeline}\label{sec:framework}
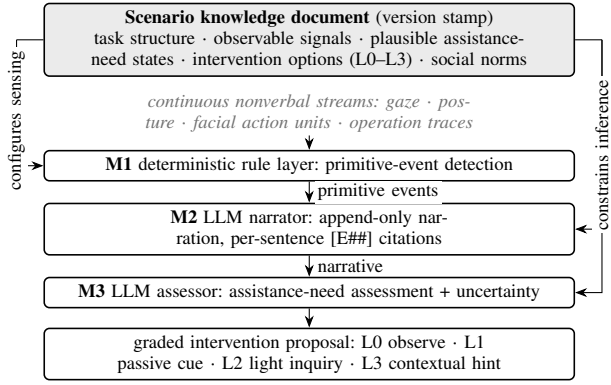
\begin{figure}[t]
\centering
\begin{tikzpicture}[
  font=\scriptsize,
  box/.style={draw, semithick, rounded corners=2pt, align=center,
              inner xsep=4pt, inner ysep=2.5pt, text width=2.65in},
  shaded/.style={box, fill=black!8},
  arr/.style={-{Stealth[length=1.6mm]}, thin},
  lab/.style={font=\scriptsize, inner sep=1pt, fill=white}
]
\node[shaded] (doc) {\textbf{Scenario knowledge document} (version stamp)\\
task structure $\cdot$ observable signals $\cdot$ plausible assistance-need
states $\cdot$ intervention options (L0--L3) $\cdot$ social norms};
\node[font=\scriptsize\itshape, text=black!55, align=center, inner sep=1pt,
      below=0.075in of doc, text width=2.5in] (sig) {continuous nonverbal
streams: gaze $\cdot$ posture $\cdot$ facial action units $\cdot$ operation traces};
\node[box, below=0.075in of sig] (m1) {\textbf{M1} deterministic rule layer:
primitive-event detection};
\node[box, below=0.11in of m1] (m2) {\textbf{M2} LLM narrator: append-only
narration, per-sentence [E\#\#] citations};
\node[box, below=0.11in of m2] (m3) {\textbf{M3} LLM assessor: assistance-need
assessment + uncertainty};
\node[box, below=0.10in of m3] (ivp) {graded intervention proposal: L0 observe
$\cdot$ L1 passive cue $\cdot$ L2 light inquiry $\cdot$ L3 contextual hint};
\draw[arr] (sig) -- (m1);
\coordinate (lrail) at ([xshift=-0.14in]doc.west);
\draw[thin] (doc.west) -- (lrail);
\draw[arr] (lrail) |- (m1.west);
\node[lab, rotate=90] at ([xshift=-0.14in, yshift=-0.42in]doc.west)
  {configures sensing};
\coordinate (rail) at ([xshift=0.14in]doc.east);
\draw[thin] (doc.east) -- (rail);
\draw[arr] (rail) |- (m2.east);
\draw[arr] (rail) |- (m3.east);
\node[lab, rotate=90] at ([xshift=0.14in, yshift=-0.6in]doc.east)
  {constrains inference};
\draw[arr] (m1) -- node[lab, right=2pt] {primitive events} (m2);
\draw[arr] (m2) -- node[lab, right=2pt] {narrative} (m3);
\draw[arr] (m3) -- (ivp);
\end{tikzpicture}
\caption{A version-controlled scenario knowledge document configures sensing
(M1) and constrains LLM reasoning (M2, M3), shaping a graded intervention
proposal. R1 is the route instantiated here; R2 (end-to-end multimodal) and R3
(gated perception) are alternatives (Section~\ref{sec:framework}).}
\label{fig:arch}
\vspace{-6pt}
\end{figure}

Fig.~\ref{fig:arch} shows the architecture. An encoder must turn continuous
nonverbal streams (gaze, posture, facial action units, operation traces, the
user's position relative to the machine) into auditable, incrementally updated
observation descriptions rather than opaque feature vectors. The document must
serve three roles at once: select which detectors run, bound what the LLM can
infer, and guide intervention choice. Perception and reasoning must run together within a live
interaction's latency budget. Separating these three concerns lets us ask, for
any new scenario, what can be reused, what reconfigured, and what must be
rebuilt.

The scenario knowledge document writes out the assumptions buried in code. Our first schema, \texttt{kiosk\_v1}, has five sections:
\emph{task structure}, \emph{observable signals}, \emph{plausible assistance-need
states}, \emph{intervention options} (L0--L3), and \emph{social norms}, plus a
version stamp (Fig.~\ref{fig:arch}). Structured sections determine
what the encoder extracts and bound what the LLM may infer; free-text norms
shape intervention choice.

Whether a new setting is an edit or a rebuild depends
on which sections change. Edits reach the LLM prompts immediately, and the encoder
swaps in a different perceptual primitive, for instance a gaze-target detector
\cite{lin2025gazehta}. Another self-service kiosk would rewrite the task structure and the social norms while keeping the
observable signals and the L0--L3 ladder. Its thresholds would need recalibrating on recordings from that setting. A setting with no screen sequence leaves little task structure standing
and needs new perceptual primitives. We have not run such a transfer, so its cost
is unmeasured.

The intervention section defines a graded ladder: \textbf{L0}, continued silent
observation; \textbf{L1}, a passive cue making help visible without
addressing the user; \textbf{L2}, a light, easily declined verbal inquiry; and
\textbf{L3}, a contextual hint referencing the current step. The
ordering follows social exposure, since a contextual hint reveals close observation of
the user's specific struggle, whereas L2 is a generic offer. Direct takeover or
staff handoff is beyond L3, outside the automated ladder by design. Choosing a
rung depends on how certain the need is and on the scenario's social norms. In
this paper we treat L0--L3 as a design space.

To keep inference inspectable, the system separates the three outputs at every
update: observation narrative, assistance-need assessment, and intervention
proposal. Kept separate, each is independently auditable, and a confident-sounding
proposal cannot hide a weakly supported observation.

Three deployment routes trade off cost, responsiveness, and auditability.
\textbf{R1} pairs an explicit lightweight encoder with an LLM reasoning over its
symbolic output. \textbf{R2} maps signals to descriptions or decisions with an
end-to-end multimodal model, possibly distilled \cite{desimone2025proactive}.
\textbf{R3}, hybrid and event-triggered, keeps costlier perception and reasoning
idle until an event triggers them. In R1 the cheap encoder runs continuously
and only the LLM calls are gated; R3 additionally gates perception itself.

This paper instantiates \textbf{R1} with multi-timescale event triggering. The
encoder is a cheap, deterministic rule layer (M1) running at sensor rate,
emitting primitive events that carry provenance to their source signals. An
event-triggered LLM narrator (M2) appends to the narrative only when new events
justify it; every sentence must cite the primitive events it summarizes (e.g.,
\texttt{[E07]}). A salience-triggered LLM assessor (M3) then reads the
frozen \texttt{kiosk\_v1} document to produce the assistance-need assessment and
an illustrative graded intervention proposal. When no salient event batch
arrives, a periodic fallback triggers it every third narration update. Reserving LLM calls for
events and salient moments, rather than captioning every frame, keeps the
pipeline responsive and its narration traceable. Section~\ref{sec:case} reports
a preliminary case study of this instantiation.

\section{Preliminary Case Study}\label{sec:case}
This preliminary case study asks whether the R1 instantiation runs the
detection-to-decision chain end to end with auditable narration. The data come from our earlier JITAS study
\cite{li2026jitas}, in which participants booked tickets at a self-service
kiosk; the harder of two task variants added friction. Synchronized
facial action units, gaze, posture, and
interface operations were recorded throughout. In a retrospective interview
afterwards, each participant rewatched the recording, marked events, and rated
struggle and willingness to receive help. We study two of these sessions, selected before any run for rich
pre-request struggle evidence and detailed retrospective accounts. Case A is dense and shorter; Case B is sparser
and far longer. Both
participants eventually requested help; our analysis window is the pre-request
phase. Sessions replay offline (not a deployed system), each event at its
recorded timestamp; LLM latencies are measured but do not gate the replay.
Timestamps count from recording start; reported durations end at
the participant's verbal help request, with sensing running briefly past it.

The scenario knowledge document was authored from the interface design, protocol,
and two non-case reference sessions, then frozen under version control before any case run.
Sessions were recorded under the original study's informed-consent protocol
\cite{li2026jitas}. Annotations and retrospective interviews were used only for
evaluation and never entered any LLM prompt. Participants are pseudonymized, and only
feature-level, de-identified text is processed. After the first runs exposed a tap-logging gap
(Section~\ref{sec:agenda}), the detector was corrected and both cases re-run once.

Only run-specific settings differ from Section~\ref{sec:framework}. The
deterministic rule layer (M1) emits nine
primitive event types under fixed, version-controlled thresholds: screen
transitions and revisits, operation stalls, long inactivity, tap bursts, scroll
thrashes, gaze aversions, posture shifts, and facial-expression-variability
spikes (stall: eight seconds idle; tap burst: four presses in two
seconds; five-second per-type refractory period). The narrator (M2) and assessor (M3)
are unchanged. The reasoning model is Claude Opus 4.8
(\texttt{claude-opus-4-8}), adaptive reasoning, no sampling-parameter overrides;
reported latencies are API-side.

\begin{table}[t]
\centering
\caption{Per-case summary (locked pipeline, frozen scenario knowledge
document). M1 = rule layer. Detection alignment =
annotated/detected/narrated (type-matched, onset within 5\,s);
fixed-cadence baseline = hypothetical 30\,s tick.}
\label{tab:cases}
\scriptsize
\setlength{\tabcolsep}{4pt}
\begin{tabular}{lrr}
\hline
 & Case A (dense) & Case B (long) \\
\hline
Time to verbal request (s) & 430 & 865 \\
M1 events (tap bursts) & 68 (14) & 103 (13) \\
Narration updates (LLM calls) & 33 & 62 \\
Uncited sentences & 0 & 0 \\
Assessor calls (salient/periodic) & 23 (22/1) & 33 (26/7) \\
Fixed-cadence baseline (30\,s) & 15 & 29 \\
Interventions L0/L1/L2/L3 & 4/7/12/0 & 3/13/16/1 \\
Need none/mon./likely/urgent & 4/0/18/1 & 2/4/26/1 \\
Median narr./assr.\ latency (s) & 4.35/20.75 & 3.93/22.41 \\
Tokens (fresh in/out) & 81{,}903/35{,}384 & 138{,}585/56{,}269 \\
Detection alignment & 13/12/12 & 7/7/7 \\
\hline
\end{tabular}
\end{table}

\begin{figure}[t]
\centerline{\includegraphics[width=3.2in]{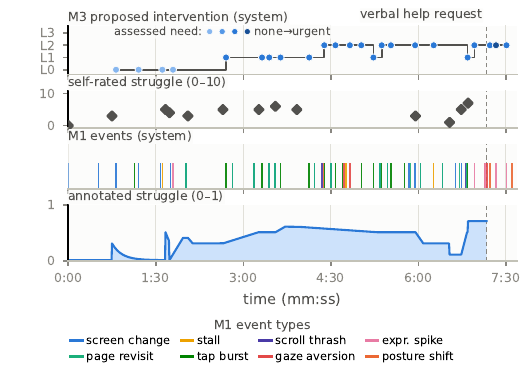}}
\caption{Case A timeline (mm:ss from recording start). Top to bottom: the M3
assessor's proposed intervention (L0--L3) as a step trace, each assessment call
a dot colored by the assessed need level (light none $\to$ dark urgent); the
participant's retrospective
self-ratings of struggle (0--10; higher = closer to giving up); M1 events,
colored by type (absent types omitted from the legend); and the struggle curve
(0--1) from the original study's annotation procedure \cite{li2026jitas}. The dashed vertical line marks the moment of the
participant's verbal help request.}
\label{fig:case}
\vspace{-6pt}
\end{figure}

Across 95 narration updates and 56 assessments there were zero uncited
sentences and no retries or empty responses.
Under Table~\ref{tab:cases}'s strict criterion, M1
detected 12 of Case A's 13 annotated episodes and all 7 of Case B's, each in
the final narrative. The zero-uncited result shows only that the narrator
followed its format; whether the cited events actually support their sentences
was not audited beyond the divergences below. Event triggering also cost more calls than a fixed 30-second tick would
have, since both sessions are dense by selection (Table~\ref{tab:cases}).\footnote{Token counts
are fresh input and output; a constant prompt prefix is additionally cached
across calls.}

Case A (Fig.~\ref{fig:case}) holds none/L0 (need level/proposed rung) through
the smooth opening minutes, turns
likely/L1 at 162~s (02:42) on a tap burst at an already-revisited screen.
From 263~s it sustains an L2 plateau that begins nearly three minutes before
the verbal request. The only \emph{urgent} in Case A (439.7~s) comes about ten seconds after
the request. The final posture, gaze, and expression cluster is the request
behavior itself. Even there the proposal stays at L2 (L3 ``would
over-intervene given remaining uncertainty'') and then drops back to likely/L1 at
session end. Because of context weighting, an earlier burst at 68~s (01:08),
just before a forward transition, is de-weighted; the 162~s burst reads as
frustration evidence.

Case B builds more slowly; none/monitor gives way to likely at 219~s, then an
L2 plateau backed by current evidence. At 798~s, 67~s before the verbal request,
the evidence converges most strongly: three tap bursts and four scroll thrashes over
the preceding minute, a posture shift and gaze aversion at 781~s (13:01)
seventeen seconds earlier. At that peak the assessor issues the only top-rung
escalation in either case, an L3 contextual hint at the document's gated ticket-amount step,
framed as ``one gradual step above the L2 the sustained struggle alone would
warrant.'' In the post-request tail, it de-escalates to monitor/L1 at 906~s as
ticket confirmation completes the booking.

The separated outputs can be checked against the participant's own retrospective
account. At
the decisive turn, all three streams coincide: the narrative's ``turning
toward the surroundings for help'' at 781~s (13:01), the assessor's sole L3 at
798~s (67~s before the request), and the participant's resolution,
``I'd pretty much exhausted everything and still couldn't get through, so I made
my peace with it: one more try, and if that failed, I'd call for help''
(translated from Japanese). All 24 participant-marked events are accounted for (author-coded,
single-coder, unblinded): 9 agree (same behavior described), 11 partial
(related, missing the specific quality), 2 diverge (contradicted), 2 out of
scope (no behavioral trace). When the participant recounts, ``Even
though I'd set the seats for the number of passengers, it still wouldn't let me
move on, and I was completely at a loss'' (translated from Japanese), the
narrative independently reports a tap-burst run on the
seat-selection screen.

The misses are as informative. Both divergences fall in Case A and share one
failure mode, a struggle-biased narrator with no positive-affect primitive. The
participant recalled two moments of relief: ``I spotted the `Information' button
and pounced on it \ldots\ and felt happy'' (translated from Japanese), then
the final advance to the next page. At both moments the narrative keeps reading ``sustained cycling
\ldots\ rather than advancing.'' The screen changes were captured, so these
failures are interpretive, not sensory. Two further
edge cases remain. A few early Case B assessments read likely while annotated struggle
was near zero (over-sensitive, though neither case reverses the rank order). Case A's one
undetected episode is an artifact of the criterion boundary: the same tapping fired six
seconds before the annotation window and the five-second refractory period
suppressed a second event.

Every reading can be traced back to its signals. The Case B sentence at 781~s cites
\texttt{[E83,E84]}, which the event log resolves to a posture-shift and a
gaze-aversion rule; M1 fired both deterministically from signal windows ending
at that timestamp, under the five-second refractory period. The interpretive sentence, the
assessment citing it, and the raw signal form one traceable chain.

\section{Discussion and Research Agenda}\label{sec:agenda}
\looseness=-1
Building the pipeline taught transferable lessons. First,
auditability came by construction. An early full-rewrite narrator drifted into
summarizing only recent events late in sessions; appending only new, cited
sentences eliminated it and left an append-only log.
Second, memory changes conclusions. With identical events and prompt, keeping
the full narrative history in the assessor's context pushed it from 11 to 18 of
21 ``likely'' verdicts
and from zero to 13 light-inquiry (L2) proposals on a non-case reference
session. Only recency weighting restored de-escalation and re-escalation as
evidence aged. Third, the audit trail proved its value. The first
runs missed every rapid-tap episode because the kiosk emits touch events the
click-based counter never saw, a gap per-event provenance exposed immediately.
After the detector fix and recalibration of the two-second tap window on
reference sessions,
alignment moved from 1/13 and 3/7 to 12/13 and 7/7. Fourth, event-triggered
cost scales with evidence density (a real cost here, a hypothesized saving on
quiet sessions), consistent with gating expensive reasoning behind lightweight
change detection \cite{deigmoeller2026merge}.

\looseness=-1
In both cases retrospective
willingness to receive help peaked well before escalation from
passive cue to light inquiry (L2), by about 100~s in Case A and 240~s in Case B. Need
is detected promptly, but escalation is conservative relative to the readiness
participants reported. This is a deliberate bias toward the norm of under-intervention in public
spaces \cite{glawe2025agency}.
Latency adds a second gap. With a median assessment latency of
twenty-one seconds, escalations at a live kiosk would arrive after the moment of
need, and real-time feasibility is untested. A preliminary check points to the time spent writing the assessment text. Published results on faster decoding and streaming output
\cite{leviathan2023specdec,kwon2023pagedattn} suggest that this cost can fall
substantially, and we will pursue that next.

\looseness=-1
Four limitations bound what two cases can show. First, the two sessions were chosen before the runs, for struggle
evidence and retrospective detail. Both contain sustained struggle, so the
pipeline was never asked to stay quiet, and specificity is untestable here. With 44 of 56 assessments at likely and two more
at urgent, the convergence claims concern timing and rank rather than
discrimination. Second, the appropriateness of the proposed interventions is not
evaluated. Third, the two cases show that the pipeline runs and can be audited.
Whether making scenario knowledge explicit works better than encoding it
implicitly is untested. Fourth, one
reasoning model, one frozen threshold set, and single-coder unblinded annotation
leave model dependence and coder reliability unknown.

\looseness=-1
The agenda covers negative-control sessions for specificity;
independent document authoring; cross-scenario
transfer, separating document replacement from encoder reconfiguration
\cite{lin2025gazehta}; a controlled comparison of the explicit document against
the same knowledge left implicit in prompts or weights; R1 versus R2 and R3
\cite{desimone2025proactive}; the
deferred appropriateness evaluation; and integration with a service robot or
cobot \cite{tanneberg2024help}.

\section{Conclusion}\label{sec:conclusion}
This paper asked what becomes possible when the scenario knowledge behind
when-and-how-to-help decisions is made explicit. In an end-to-end two-case
study, a single version-controlled document configured sensing, constrained
LLM reasoning, and kept every inference step auditable. The narration stayed
traceable to its primitive events, and the assessor followed two different
struggle dynamics with graded restraint. The appropriateness evaluation with users comes next. Beyond it lie
sessions without struggle, transfer to another scenario, and a robot that acts on
the proposals.

\bibliographystyle{IEEEtran}

\end{document}